\documentclass[aps,prb,showpacs,twocolumn,superscriptaddress]{revtex4}
\usepackage{amsmath,graphics}
\usepackage[next]{inputenc}
\usepackage{graphicx}
\usepackage{hyperref}
\def\be{\begin{equation}}
\def\ee{\end{equation}}

\def\bi{\begin{itemize}}
\def\ei{\end{itemize}}
\def\bn{\begin{enumerate}}
\def\en{\end{enumerate}}
\def\bea{\begin{eqnarray}}
\def\eea{\end{eqnarray}}
\def\no{\nonumber}
\def\ba{\begin{array}}
\def\ea{\end{array}}
\def\bd{\begin{displaymath}}
\def\ed{\end{displaymath}}

\begin{document}

\title{Thermodynamic Properties of the One-Dimensional Extended Quantum Compass Model
in the Presence of a Transverse Field}
\author{R. Jafari}
\affiliation{Research Department, Nanosolar System Company (NSS), Zanjan 45158-65911, Iran}
\email[]{jafari@iasbs.ac.ir, jafari@nss.co.ir}
\affiliation{Department of Physics, Institute for Advanced
Studies in Basic Sciences (IASBS), Zanjan 45137-66731, Iran}

\begin{abstract}
The presence of a quantum critical point can significantly affect the
thermodynamic properties of a material at finite temperatures. This is
reflected, \textit{e.g.}, in the entropy landscape $S(T, c)$ in the
vicinity of a quantum critical point, yielding particularly strong
variations for varying the tuning parameter $c$ such as magnetic field.
In this work we have studied the thermodynamic properties of the quantum
compass model in the presence of a transverse field. The specific heat,
entropy and cooling rate under an adiabatic demagnetization process have
been calculated. During an adiabatic (de)magnetization process temperature
drops in the vicinity of a field-induced zero-temperature quantum phase transitions.
However close to field-induced quantum phase transitions we observe a large
magnetocaloric effect.
\end{abstract}
\date{\today}


\maketitle

%%%%%%%%%%%%%%%%%%%%%%%%%%%%%%%%%%%%%%%%%%%%%%%%%%%%%%%%%%%%%%%%%%%%%
\section{Introduction \label{introduction}}

The magnetocaloric effect (MCE), in general, addresses the change of
temperature of magnetic systems under the adiabatic variation of an external
magnetic field which was discovered in iron by Warburg in 1881 \cite{Warburg}.
The MCE has been widely used for refrigeration due to potential room-temperature
cooling applications \cite{Gschneider,Tishin}. For example, adiabatic demagnetization
of paramagnetic salts was the first method to reach temperatures below 1K \cite{MacDougall}
whereas demagnetization of nuclear spins has reached temperatures down to 100pK
\cite{Oja} and is still the method of choice for $\mu$K-range cooling \cite{Strehlow}.
The cooling rate at the adiabatic demagnetization
$(\frac{\partial T}{\partial h})$ for an ideal paramagnet (i.e., a system of
non-interacting magnetic moments) is equal to $T/H$, which means linear monotonic
dependence of temperature on strength of the magnetic field. This linear dependence
rise out of a direct consequence of the fact that for any paramagnetic system the
entropy depends only on the ratio $T/H$, so for any isentrope one gets $H/T$ = const.
However, the matter could undergo crucial changes in systems with interacting spins.
For instance, in ferromagnets near the Curie point one can observe a substantial enhancement
of the effect \cite{Tishin}.

As has been shown in early investigations, quantum antiferromagnets are more
efficient low-temperature magnetic coolers than ferromagnets \cite{Bonner}.
This fact is connected with the behavior of the entropy of antiferromagnets.
The entropy of any antiferromagnet at low temperatures displays (at least) one
maximum as a function of magnetic field, which usually, according to the third
law of thermodynamics, falls to zero at $T\rightarrow 0$.

However, the MCE in quantum spin systems has recently attracted scientists attention.
From one hand, field-induced quantum phase transitions lead
to universal responses when the applied field is varied adiabatically \cite{Zhu,Garst,Honecker1,Zhitomirsky1}.
On the other hand, it was observed that the MCE is enhanced by geometric frustration
\cite{Zhitomirsky2,Honecker2,Schmidt,Pereira}, promising improved efficiency in
low-temperature cooling applications. More generally, the MCE is particularly large in the
vicinity of quantum critical points (QCPs). The MCE is closely related to the generalized Gr\"{u}neisen ratios
\bea
\no
\Gamma_{c}=-\frac{1}{T}\frac{(\partial S/\partial C)_{T}}{(\partial S/\partial T)_{c}},
\eea
where $c$ is the control parameter governing the quantum phase transition 
which would be external magnetic field $h$ for MCE. Using basic
thermodynamic relations \cite{Tishin}, the generalized Gr\"{u}neisen ratio
$\Gamma_{h}$ can be related to the adiabatic cooling rate $(\partial T/\partial H)_{S}$ \cite{Trippe}
\bea
\no
\Gamma_{h}=\frac{1}{T}\Big(\frac{\partial T}{\partial h}\Big)_{S}=
-\frac{1}{C_{c}}\Big(\frac{\partial M}{\partial T}\Big)_{h}.
\eea
The magnetic cooling rate is an important quantity for the characterization
of quantum critical points (QCPs), \textit{i.e.}, quantum phase transitions between different magnetic structures
under tuning the magnetic field at $T=0$.

Quantum Compass Model (QCM) is simplified model which describes the
nature of the orbital states in the case of a twofold degeneracy \cite{Kugel}. Originally, the QCM
has been used to describe the Mott insulators with orbit degeneracies.
It depends on the lattice geometry and belongs to the low energy Hamiltonian originated from the
magnetic interactions in Mott-Hubbard systems with the strong spin-orbit coupling \cite{Jackeli}.
In QCM the orbital degrees of freedom are represented by pseudospin operators and
coupled anisotropically in such a way as to mimic the competition between orbital
orderings in different directions. For simplicity, the one-dimensional (1D) QCM, is
constructed by antiferromagnetic order of $X$ and $Y$ pseudospin components on odd and even bonds, respectively
\cite{Brzezicki,Brzezicki2}.
Moreover, the extended version of the 1D QCM, is obtained by introducing one more tunable parameter,
has been studied by Eriksson \textit{et al}. \cite{Eriksson}.

To the best of our knowledge, thermodynamic properties of the extended version of 1D QCM in a transverse field
has not been studied so far. Since the extended QCM shows two critical lines (first order and second order 
transition lines), investigation on its thermodynamic properties and MCE will be of great importance.
However, in our recent work, we have shown that extended QCM in a transverse field reveals
a rich phase diagram which includes several critical surfaces depending on exchange couplings \cite{Jafari}.
In this paper we exploit the method employed in Refs. [\onlinecite{Brzezicki2}] and [\onlinecite{Jafari}]
to investigate  thermodynamic properties of extended QCM in absent and presence of a transverse magnetic field.

\section{Hamiltonian and Exact Solution\label{EQCMTF}}
Consider the Hamiltonian
\bea
\no
H=\sum_{n=1}^{N'}&[&J_{1}\sigma^{x}_{2n-1}\sigma^{x}_{2n}+
J_{2}\sigma^{y}_{2n-1}\sigma^{y}_{2n}+ L_{1}\sigma^{x}_{2n}\sigma^{x}_{2n+1}\\
\label{eq1}
&+&h_{1}\sigma^{z}_{2n-1}+h_{2}\sigma^{z}_{2n}].
\eea
where $J_{1}$ and $J_{2}$ are the odd bonds exchange couplings, $L_{1}$ is the even bond exchange coupling
and the parameters $h_{1}$ and $h_{2}$ describe the external magnetic field at the even and odd
lattice sites respectively which its alternating nature is assumed to arise from different magnetic 
moments or different g-factors of the spin.
The number of sites is $N=2N'$ and for simplicity we assume periodic boundary conditions.
The above Hamiltonian (Eq. (\ref{eq1})) can be exactly diagonalized by standard Jordan-Wigner
transformation \cite{Jordan,Perk} as defined below,

\bea
\no
\sigma^{x}_{j}=b^{+}_{j}+b^{-}_{j},~~
\sigma^{y}_{j}=b^{+}_{j}-b^{-}_{j},~~
\sigma^{z}_{j}=2b^{+}_{j}b^{-}_{j}-1
\eea
\bea
\no
b^{+}_{j}=c^{\dag}_{j}~e^{i\pi\Sigma_{m=1}^{j-1}c^{\dag}_{m}c_{m}},~~
b^{-}_{j}=e^{-i\pi\Sigma_{m=1}^{j-1}c^{\dag}_{m}c_{m}}~c_{j}
\eea

which transforms spins into fermionic operators $c_{j}$.

The crucial step is to define independent Majorana fermions \cite{Sengupta} at site $n$,
$c_{n}^{q}\equiv c_{2n-1}$ and $c_{n}^{p}\equiv c_{2n}$. This can be regarded as quasiparticles' spin
or as splitting the chain into bi-atomic
elementary cells \cite{Brzezicki2}.

Substituting for $\sigma^{x}_{j}$, $\sigma^{y}_{j}$ and $\sigma^{z}_{j}$ ($j=2n, 2n-1$) in terms of
Majorana fermions with antiperiodic boundary condition (subspace with even number of fermions) followed by a
Fourier transformation, Hamiltonian Eq. (\ref{eq1}) (apart from additive constant), can be written as

\bea
\no
H^{+}=\sum_{k}\Big[Jc_{k}^{q\dag}c_{-k}^{p\dag}+Lc_{k}^{q\dag}c_{k}^{p}
+2h_{1}c_{k}^{q\dag}c_{k}^{q}+2h_{2}c_{k}^{p\dag}c_{k}^{p}+h.c.\Big],
\eea

where $J=(J_{1}-J_{2})-L_{1}e^{ik}$, $L=(J_{1}+J_{2})+L_{1}e^{ik}$ and $k=\pm\frac{j\pi}{N'},~(j=1,3,\cdots,N'-1)$.\\

It should be pointed out that though the GS in periodic and antiperiodic boundary conditions are slightly
different in the finite-size system, they are identical in the thermodynamic limit and the essential features
in finite size are also not altered qualitatively.

Finally, diagonalization is completed by a four-dimensional Bogoliubov transformation connecting
$c_{k}^{q\dag},~c_{-k}^{q},~c_{k}^{p\dag},~c_{-k}^{p}$ and one thus obtains two different
kind of quasiparticles,

\bea
\label{eq2}
H=\sum_{k}\Big[E^{q}_{k}(\gamma_{k}^{q\dag}\gamma_{k}^{q}-\frac{1}{2})+
E^{p}_{k}(\gamma_{k}^{p\dag}\gamma_{k}^{p}-\frac{1}{2})\Big],
\eea

where $E^{q}_{k}=\sqrt{2(a+c)}$ and $E^{p}_{k}=\sqrt{2(a-c)}$, $c=\sqrt{a^{2}-b}$ in which

\bea
\no
a&=&h_{1}^{2}+h_{2}^{2}+J_{1}^{2}+J_{2}^{2}+L_{1}^{2}+2L_{1}J_{2}\cos k,\\
\no
b&=&4[(J_{1}J_{2}-h_{1}h_{2})^{2}+2J_{1}L_{2}(J_{1}J_{2}-h_{1}h_{2})\cos k+J_{1}^{2}L_{1}^{2}].
\eea

The ground state ($E_{G}$) and the lowest excited state ($E_{E}$) energies are obtained from Eq.(\ref{eq3}),

\bea
\no
E_{G}=-\frac{1}{2}\sum_{k}(E^{q}_{k}+E^{p}_{k}),~~E_{E}=-\frac{1}{2}\sum_{k}(E^{q}_{k}-E^{p}_{k}),
\eea

which can be written as a function of $a$ and $b$,

\bea
\label{eq3}
E_{G}=-2\sum_{k>0}\sqrt{a+\sqrt{b}},~~E_{E}=-2\sum_{k>0}\sqrt{a-\sqrt{b}}
\eea

It is clear that the ground state is separated from the lowest energy pseudospin excitation by a pseudospin
gap $\Delta=|E_{E}-E_{G}|$,	which vanishes at $h_{1}h_{2}=J_{1}(J_{2}\pm L_{1})$ in the thermodynamic limit.

It should be mentioned that the exact spectrum and the pseudospin
gap are the same as that obtained in our recent work for $h_{1}=h_{2}=h$ \cite{Jafari}.

For investigation on the thermodynamic properties of the model, we have calculated the free
energy per sit \cite{}.

\bea
\label{eq4}
f=-\frac{1}{\beta}\lim_{N\rightarrow\infty}\big[\frac{1}{N}\ln Sp\exp(-\beta H)\big],
\eea

where $Sp$ means a summation over all occupation patterns. The free energy is easily obtained by using the
diagonalized quadratic form of $H$ (Eq.(\ref{eq2})),

\begin{widetext}
\bea
\label{eq5}
f=-\frac{1}{\beta}\frac{1}{2\pi}\int_{-\pi}^{\pi}\ln\Big[2\cosh\big(\frac{\beta(E^{q}_{k}+E^{p}_{k})}{2}\big)
+2\cosh\big(\frac{\beta(E^{q}_{k}-E^{p}_{k})}{2}\big)\Big]dk
\eea
\end{widetext}

Entropy ($S$), specific heat ($C_{v}$) and the normalized cooling rate ($\Gamma_{c}$) are related to
free energy via a simple thermodynamic relation,

\bea
\no
S=\beta^{2}\frac{\partial f}{\partial\beta},~~~C_{v}=-\beta\frac{\partial S}{\partial\beta},~~~
\Gamma_{c}=-\frac{\partial^{2}f}{\partial T\partial c}/(T\frac{\partial^{2}f}{\partial T^{2}}).
\eea

Then, from the Eq.(\ref{eq5}) one can easily obtain simple analytic expressions for all thermodynamic
functions of the system. The adiabatic demagnetization curves can be found from a direct solution of
$S(c,T)=const$.

%%%%%%%%%%%%%%%%%%%%%%  Fig.1   %%%%%%%%%%%%%%%%%%%%%%%
\begin{figure}[b]
\begin{center}
\includegraphics[width=9.3cm]{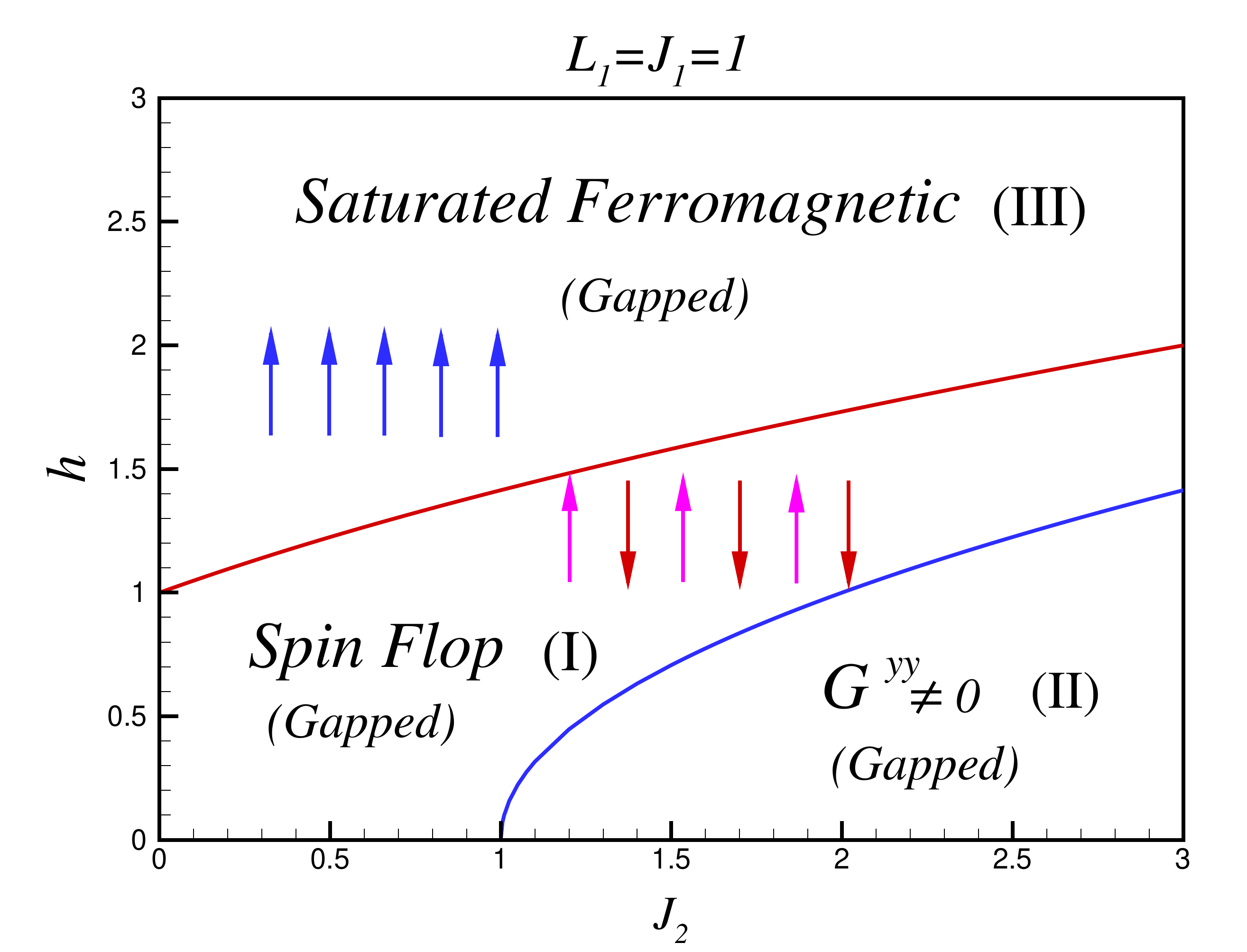}
\caption{(Color online) Zero-temperature phase diagram of the
extended quantum compass model in the transverse magnetic field
for $L_{1}=J_{1}=1$.} \label{fig1}
\end{center}
\end{figure}
%%%%%%%%%%%%%%%%%%%%%%%%%%%%%%%%%%%%%%%%%%%%%%%%%%%%%%%

%%%%%%%%%%%%%%%%%%%%%%  Fig.2   %%%%%%%%%%%%%%%%%%%%%%%
\begin{figure}[b]
\begin{center}
\includegraphics[width=9cm]{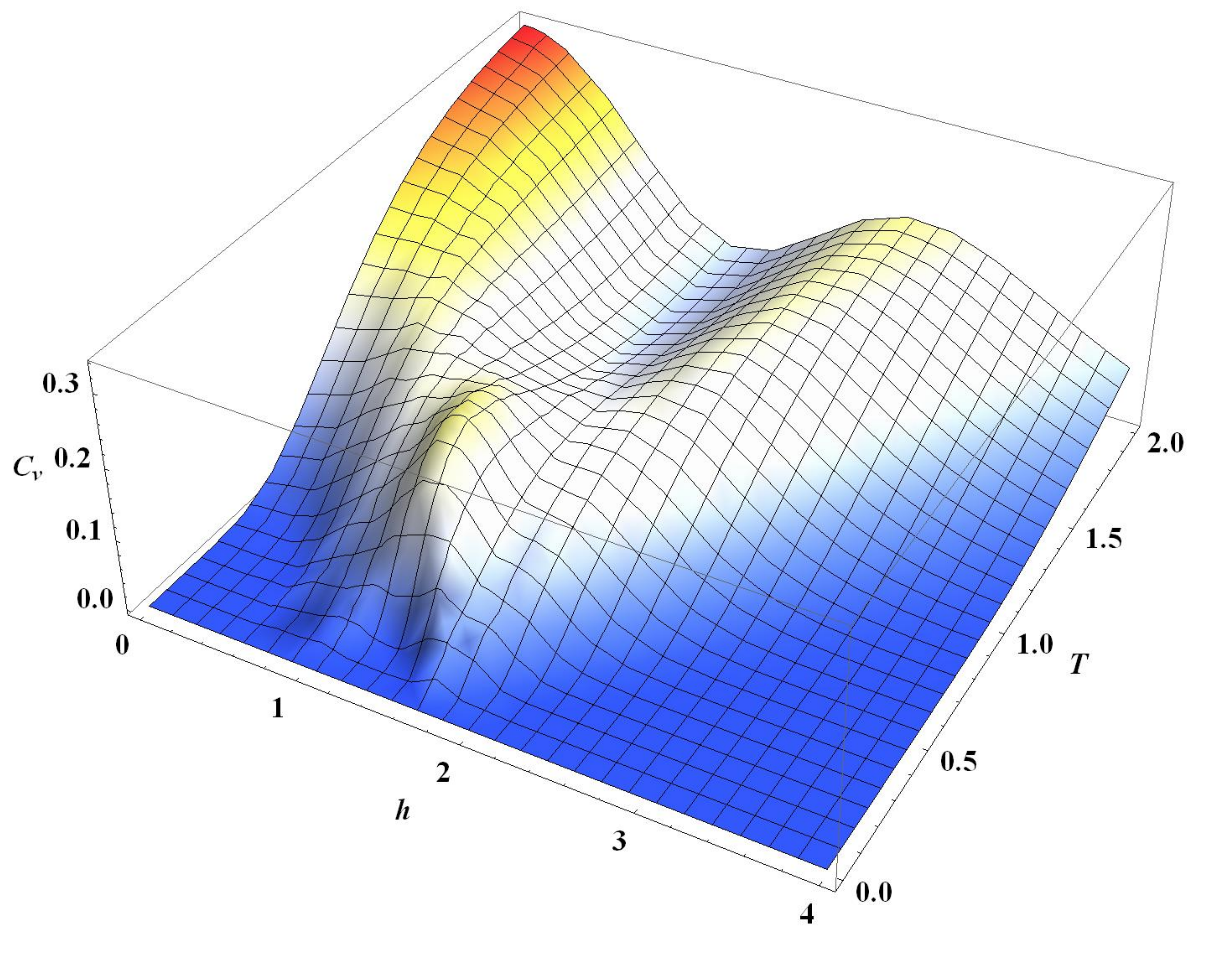}
\caption{(Color online) The three-dimensional plot of
specific heat of extended quantum compass model in a transverse
field versus temperature and magnetic field for $J_{2}=2$. For
extremely low temperature the maximums of the specific heat happen
at $h_{c_{1}}=1$ and $h_{c_{2}}=\sqrt{3}$.} \label{fig2}
\end{center}
\end{figure}
%%%%%%%%%%%%%%%%%%%%%%%%%%%%%%%%%%%%%%%%%%%%%%%%%%%%%%%

\section{Extended Quantum Compass Model in a Homogenous Magnetic Field ($h_{1}=h_{2}=h$)\label{ED}}

In our recent work, we have investigated the phase diagram of the extended quantum compass model in
homogenous transverse field by use of the gap analysis and universality of
derivative of the correlation functions (ground state) \cite{Jafari}. This model is always gapful except
at the critical surfaces where the energy gap disappears. We have obtained the analytic
expressions for all critical fields which drive quantum phase transitions (QPT) as a function
of exchange couplings, $h_{0}=\sqrt{J_{1}(J_{2}+L_{1})}$, $h_{\pi}=\sqrt{J_{1}(J_{2}-L_{1})}$.

%%%%%%%%%%%%%%%%%%%%%%  Fig.3   %%%%%%%%%%%%%%%%%%%%%%%
\begin{figure}[t]
\begin{center}
\includegraphics[width=9cm]{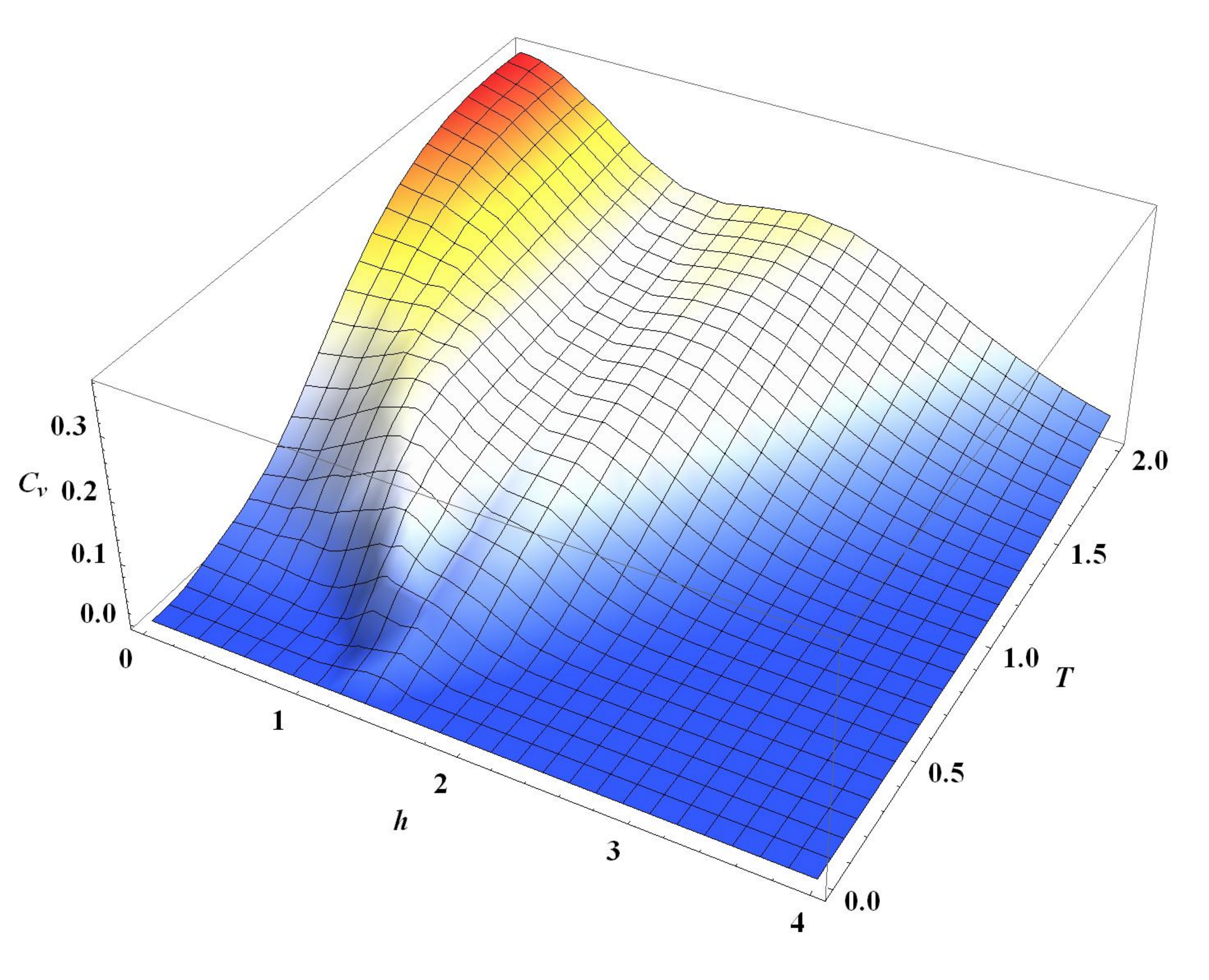}
\caption{(Color online) The specific heat of extended quantum
compass model in a transverse field versus temperature and magnetic
field for $J_{2}=0.8$. In this case there is only one maximum at
$h_{c}=\sqrt{1.8}$ for very low temperature.} \label{fig3}
\end{center}
\end{figure}
%%%%%%%%%%%%%%%%%%%%%%%%%%%%%%%%%%%%%%%%%%%%%%%%%%%%%%%

%%%%%%%%%%%%%%%%%%%%%  Fig.4   %%%%%%%%%%%%%%%%%%%%%%%
\begin{figure}
\begin{center}
\includegraphics[width=9cm]{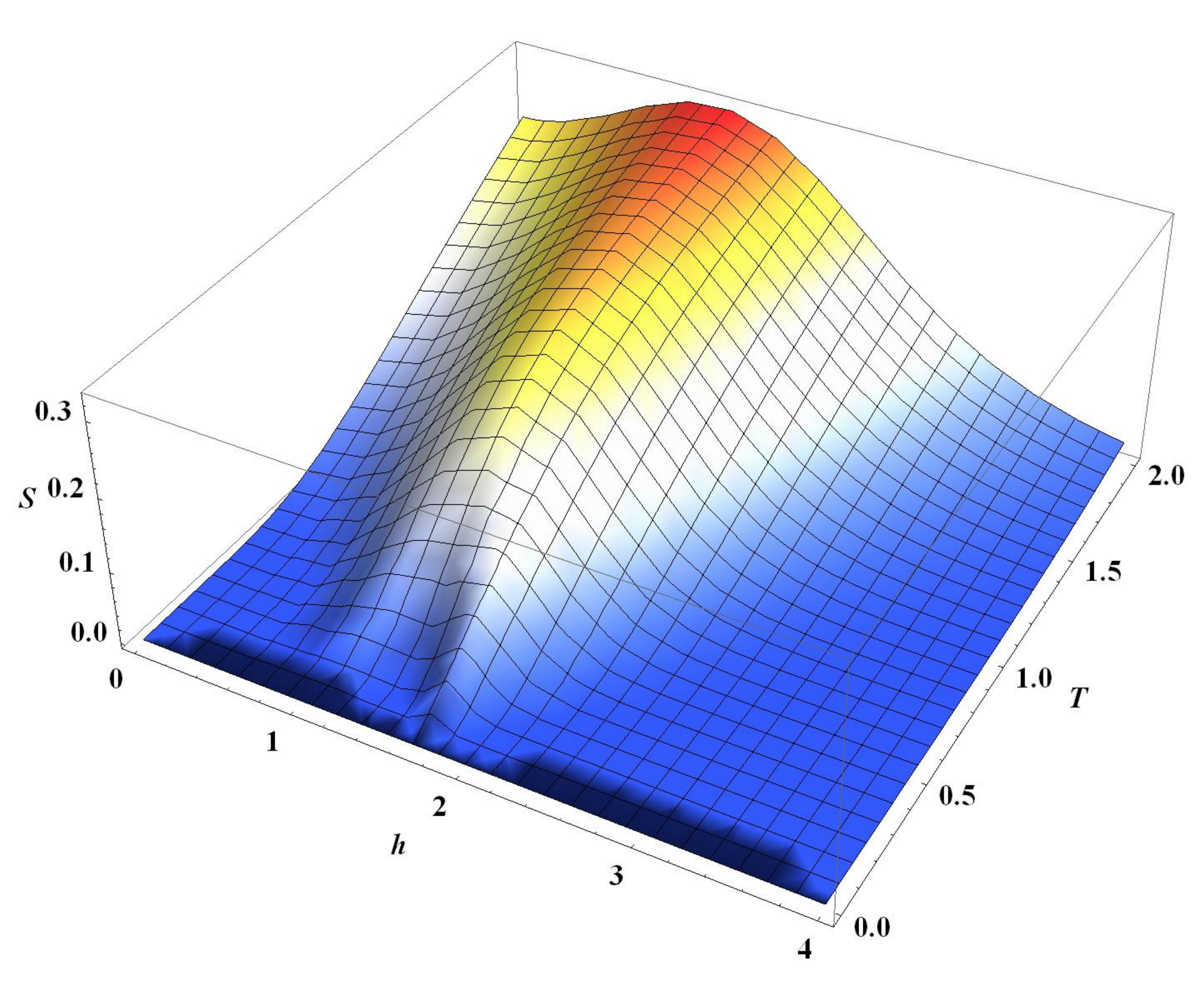}
\caption{(Color online) The three-dimensional
panorama of entropy of extended quantum compass model in a transverse field
versus temperature and magnetic field for $J_{2}=2$. The zero-temperature
critical fields are signaled by maximum of the entropy at $h_{c_{1}}=1$ and $h_{c_{2}}=\sqrt{3}$
for low temperature.} \label{fig4}
\end{center}
\end{figure}
%%%%%%%%%%%%%%%%%%%%%%%%%%%%%%%%%%%%%%%%%%%%%%%%%%%%%%%

%%%%%%%%%%%%%%%%%%%%%%  Fig.5   %%%%%%%%%%%%%%%%%%%%%%%
\begin{figure}[t]
\begin{center}
\includegraphics[width=9cm]{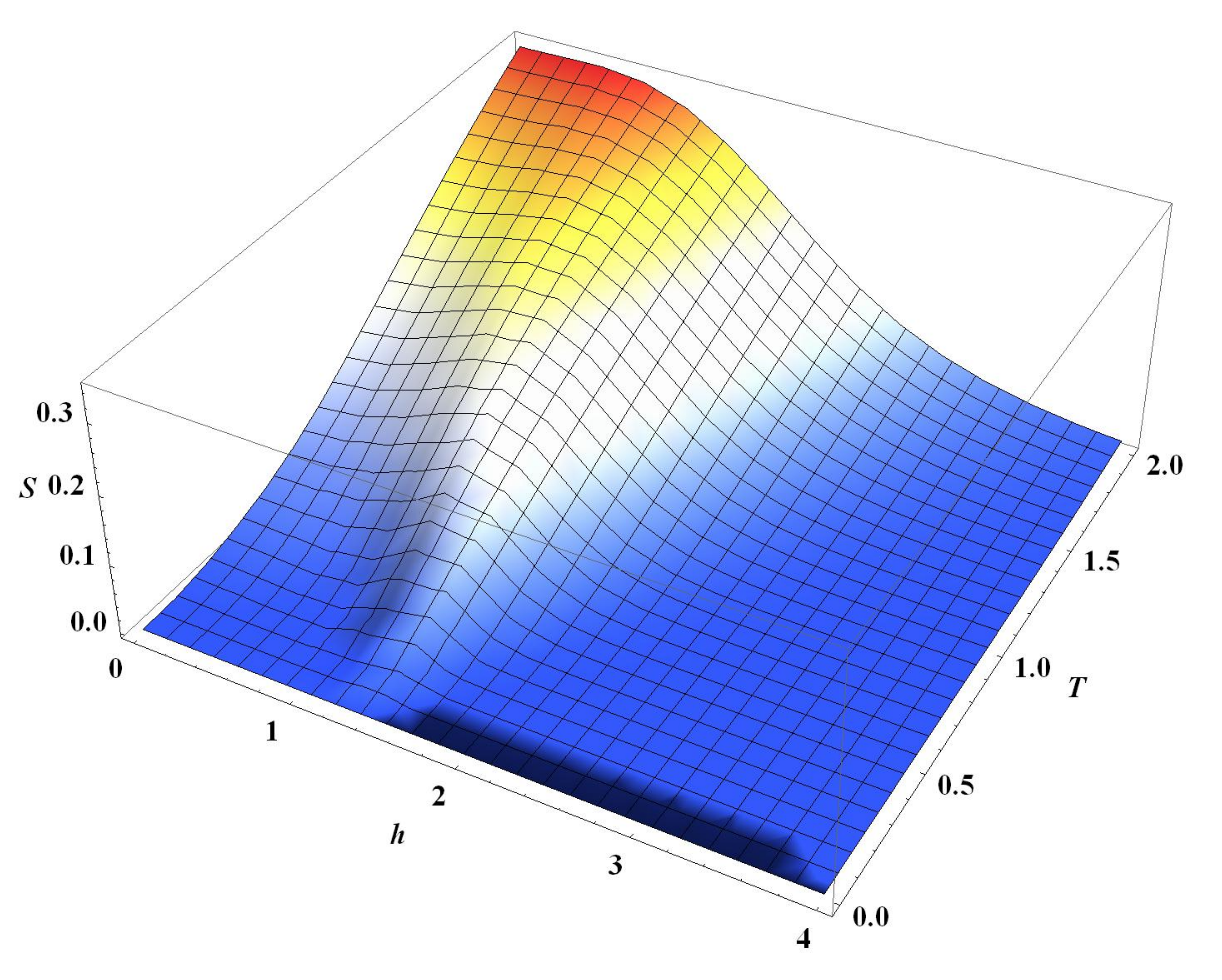}
\caption{(Color online) The entropy of extended quantum compass model
in a transverse field versus temperature and magnetic field for $J_{2}=0.8$.
In this case the zero-temperature critical field ($h_{c}=\sqrt{1.8}$)
is specified by maximum of entropy for extremely low temperature.} \label{fig5}
\end{center}
\end{figure}
%%%%%%%%%%%%%%%%%%%%%%%%%%%%%%%%%%%%%%%%%%%%%%%%%%%%%%%

It is useful to recall the zero-temperature phase diagram of the extended compass model
in homogenous transverse field, see Fig. (\ref{fig1}) \cite{Jafari} (For simplicity we take $J_{1}=L_{1}=1$).
For $J_{2}<1$ and small magnetic field $h<h_{0}$ (region (I)) the model is in the
spin-flop phase (the Ne\'{e}l order along the axis ($x$) which is perpendicular to magnetic
field is called spin flop). In this region tuning the magnetic field forces the system to
fall into a saturated ferromagnetic (SF) phase above the critical field ($h_{c}=h_{0}$). 
These states exhibit a gap in the excitation spectrum where vanishes at the critical field.
In the case of $J_{2}>1$ there is antiparallel ordering of spin y component on odd bonds
under the lower critical field $h_{c_{1}}=h_{\pi}$ (region (II)) and beyond this critical
field system goes into a gapped spin-flop phase (region (I)). This spin-flop phase exist
for $h<h_{c_{2}}=h_{0}$. Finally, for $h>h_{c_{2}}$ the ground state is the ferromagnetically
polarized state along the magnetic field (region (III)) which exhibits again a gap.
These different zero-temperature regions, in particular the quantum phase transitions
at $h_{0}$ and $h_{\pi}$ could be reflected by the magnetocaloric properties at finite
temperature \cite{Honecker1,Zhitomirsky1,Zhitomirsky2,Honecker2}.

%%%%%%%%%%%%%%%%%%%%%%  Fig.6   %%%%%%%%%%%%%%%%%%%%%%%
\begin{figure}[t]
\begin{center}
\includegraphics[width=9.cm]{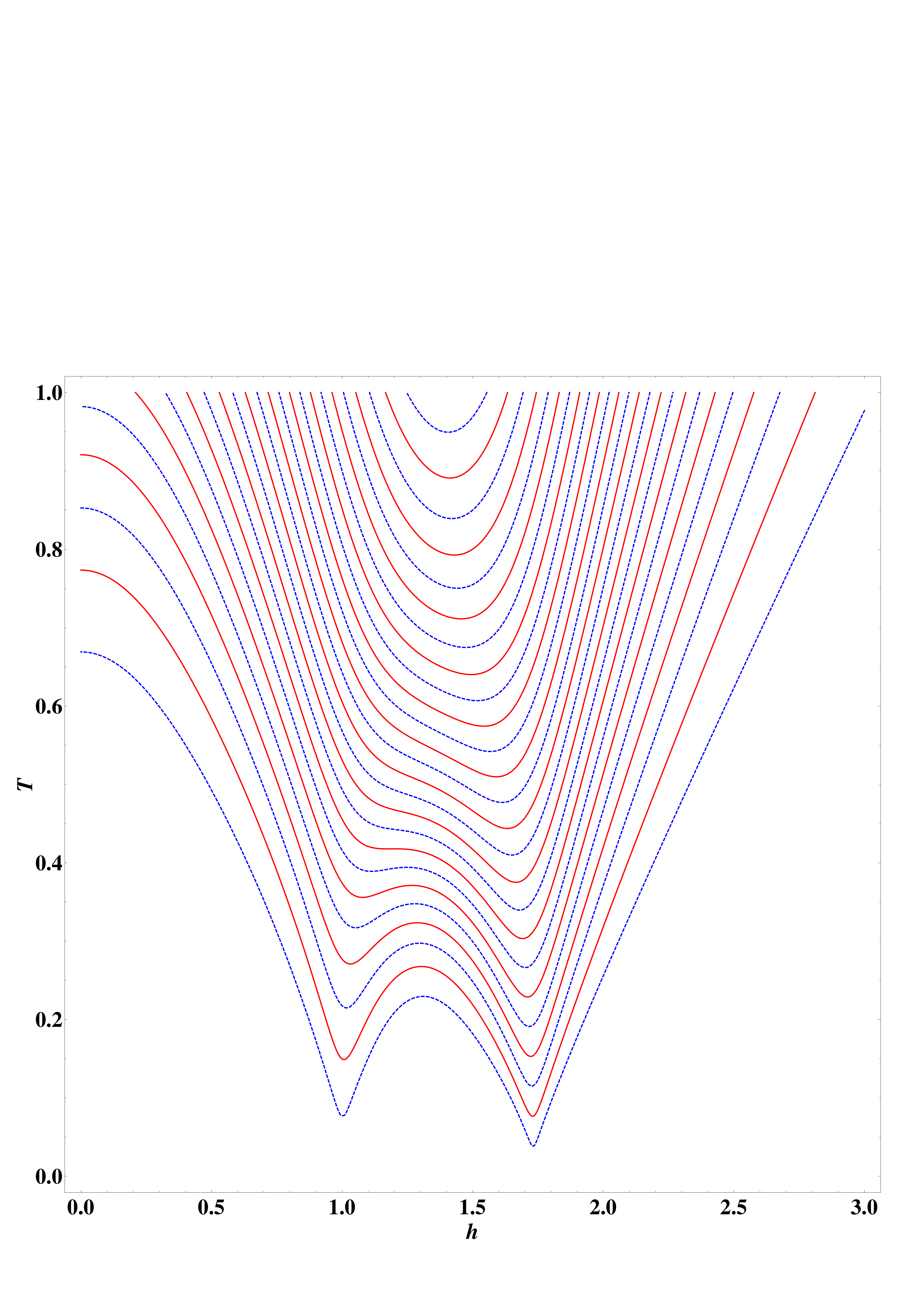}
\caption{(Color online) Adiabatic demagnetization curves
of the extended quantum compass model in a transverse field
for $J_{2}=2$. The lowest temperatures occur at zero-temperature
critical fields $h_{c_{1}}=1$ and $h_{c_{2}}=\sqrt{3}$.} \label{fig6}
\end{center}
\end{figure}
%%%%%%%%%%%%%%%%%%%%%%%%%%%%%%%%%%%%%%%%%%%%%%%%%%%%%%%

%%%%%%%%%%%%%%%%%%%%%%  Fig.7   %%%%%%%%%%%%%%%%%%%%%%%
\begin{figure}
\begin{center}
\includegraphics[width=9cm]{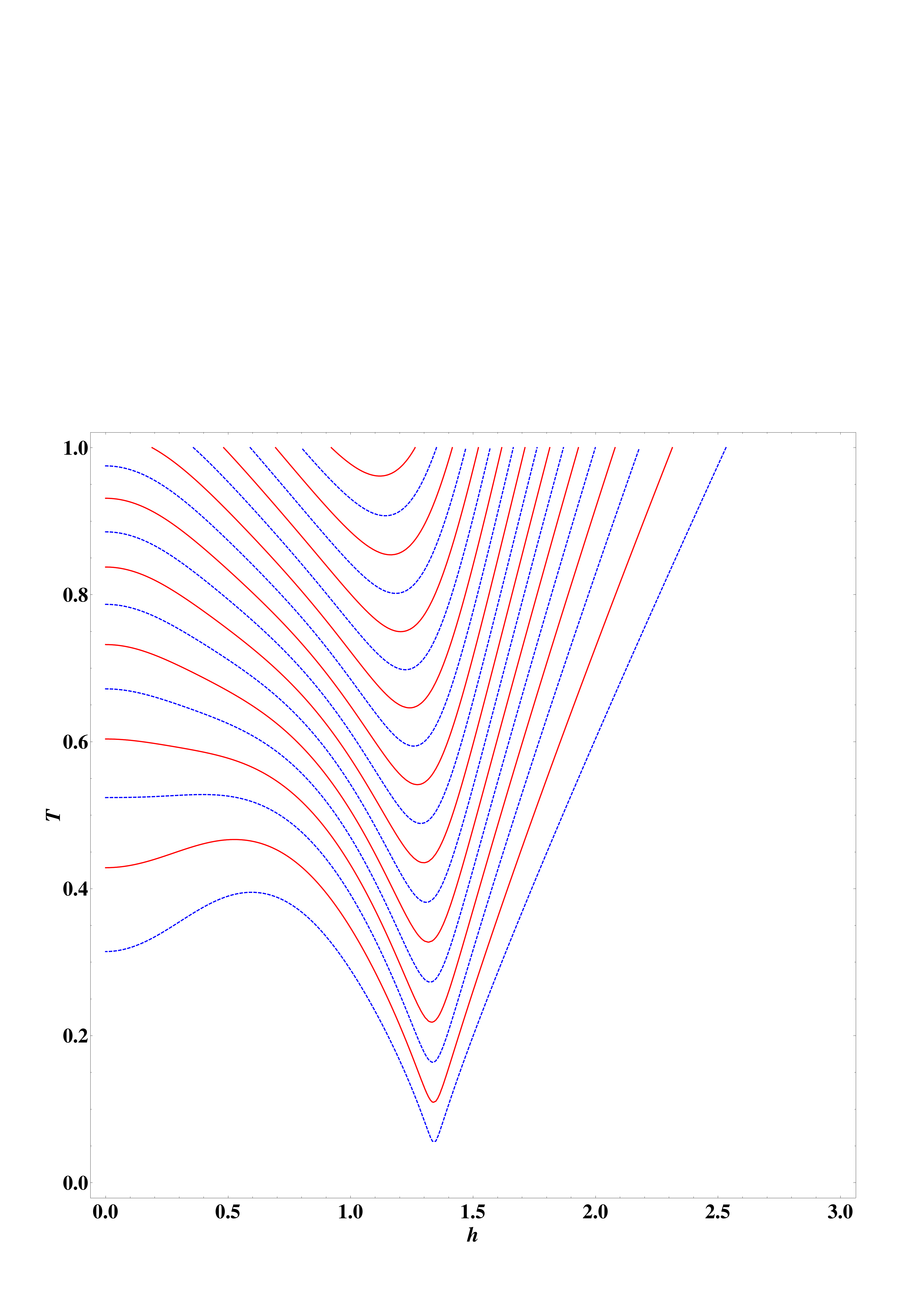}
\caption{(Color online) Lines of constant entropy, \textit{i.e.},
adiabatic demagnetization curves of the extended quantum compass model
in a transverse field for $J_{2}=0.8$. The minimum of isentrop locates at
the zero-temperature critical field $h_{c}=\sqrt{1.8}$} \label{fig7}
\end{center}
\end{figure}
%%%%%%%%%%%%%%%%%%%%%%%%%%%%%%%%%%%%%%%%%%%%%%%%%%%%%%%

The specific heat of extended quantum compass model in transverse field has been
plotted in Figs. (\ref{fig2}) and (\ref{fig3}) versus temperature ($T$) and magnetic
field ($h$) for $J_{2}=2$ and $J_{2}=0.8$ respectively.
As it can be seen from Fig. (\ref{fig2}), the specific heat reaches its maximums at QCP for extremely
low temperatures producing the a small well between two critical fields $h_{c_{1}}=h_{\pi}=1$
and $h_{c_{2}}=h_{0}=\sqrt{3}$, while there is only one maximum at the
critical field $h_{c}=h_{0}=\sqrt{1.8}$ in Fig. (\ref{fig3}).

Figs (\ref{fig4}) and (\ref{fig5}) show the entropy of the model for
$J_{2}=2$ and $J_{2}=0.8$. They manifest that the maximums of the entropy for very
low temperature occur at the critical fields just like the specific heat. This accumulation of entropy
close QCPs indicates that the system is maximally undecided which ground state to chose.
As the temperature is raised, all the characteristic behaviors have been disappeared.

%%%%%%%%%%%%%%%%%%%%%%  Fig.8   %%%%%%%%%%%%%%%%%%%%%%%
\begin{figure}
\begin{center}
\includegraphics[width=9cm]{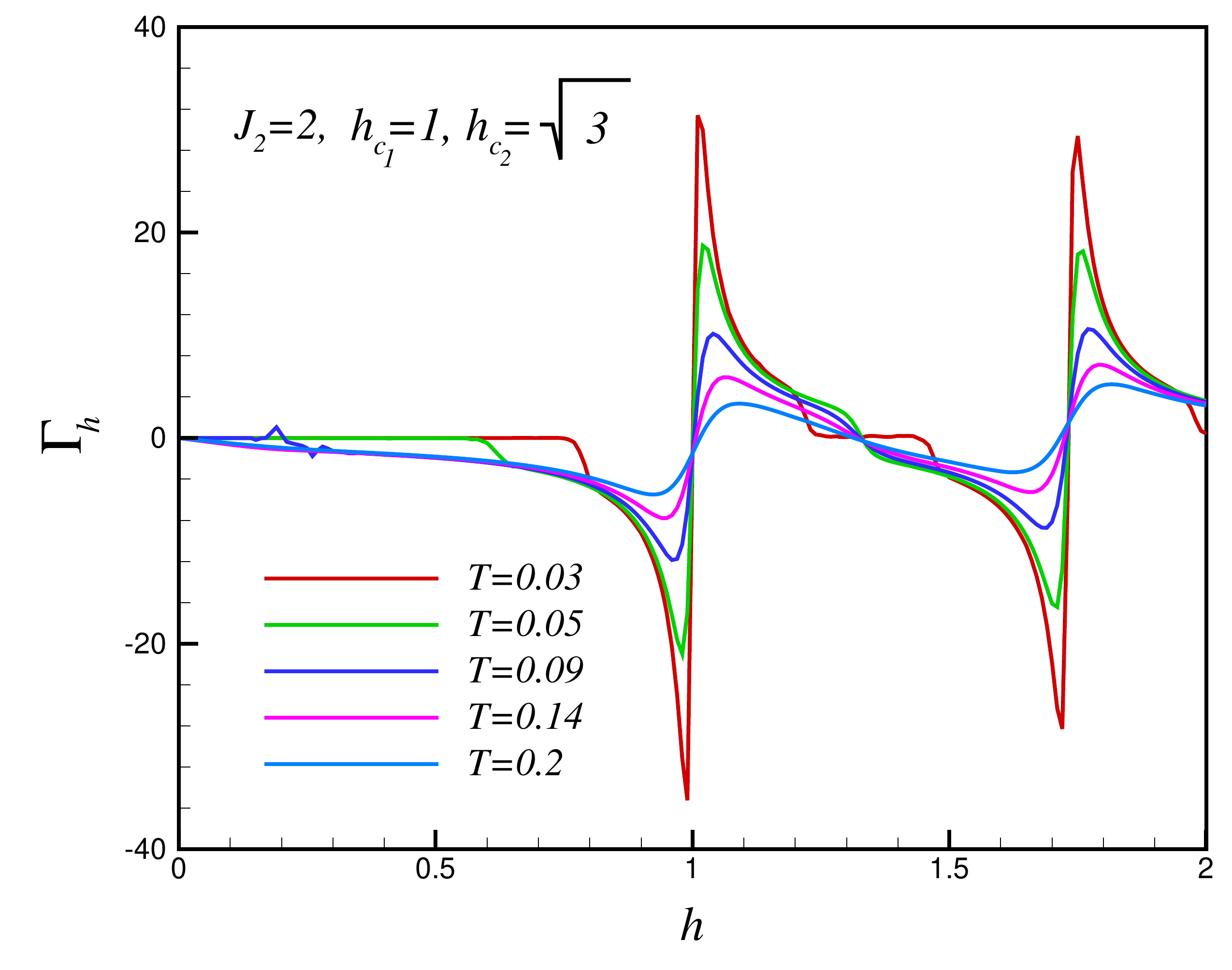}
\caption{(Color online) Magnetic cooling rate ($\Gamma_{h}$) versus
magnetic field for different values of temperature for $J_{2}=2$.} \label{fig8}
\end{center}
\end{figure}
%%%%%%%%%%%%%%%%%%%%%%%%%%%%%%%%%%%%%%%%%%%%%%%%%%%%%%%

%%%%%%%%%%%%%%%%%%%%%%  Fig.9   %%%%%%%%%%%%%%%%%%%%%%%
\begin{figure}
\begin{center}
\includegraphics[width=9cm]{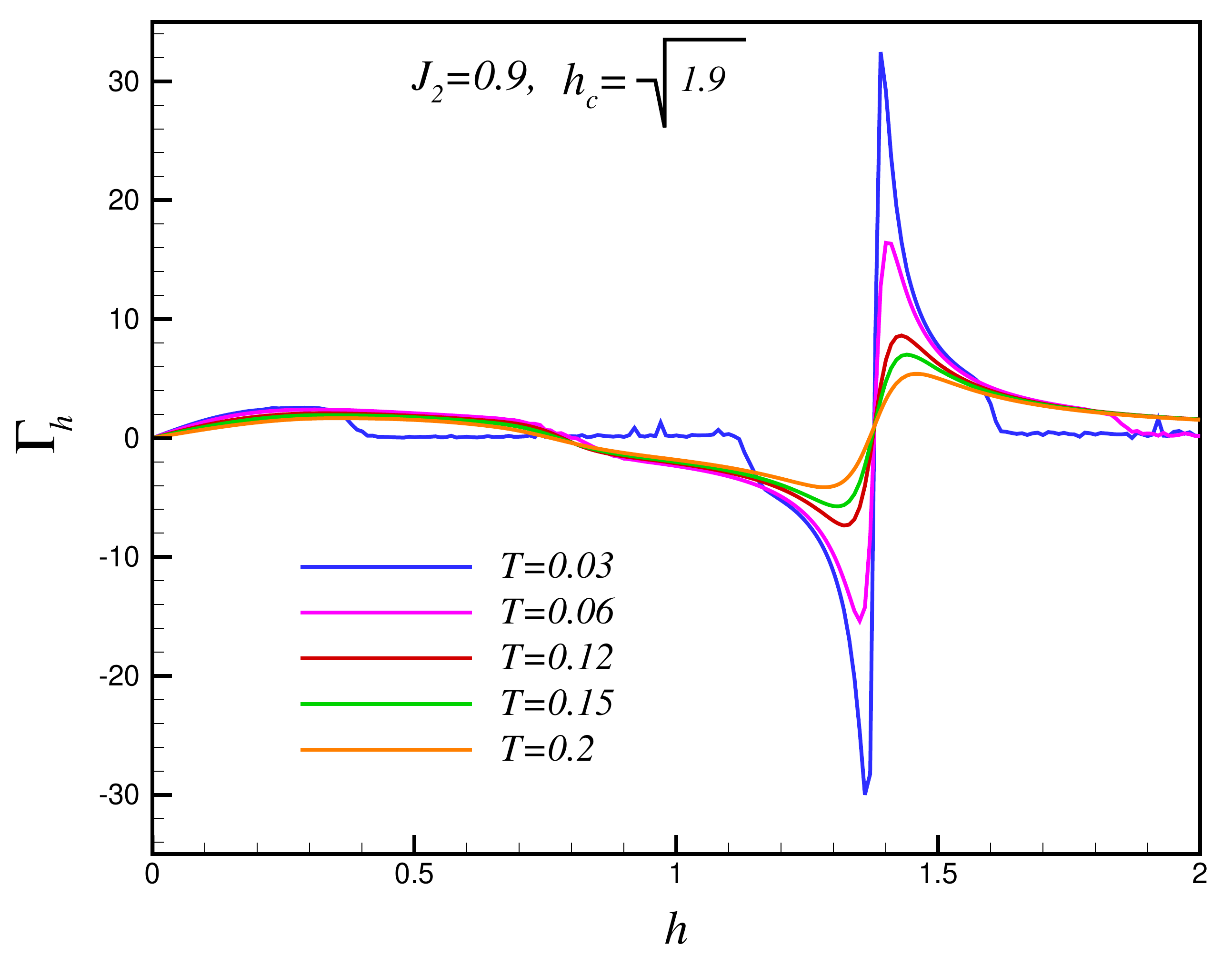}
\caption{(Color online) The variations of cooling rate ($\Gamma_{h}$) with
magnetic field for different values of temperature for $J_{2}=0.9$.} \label{fig9}
\end{center}
\end{figure}
%%%%%%%%%%%%%%%%%%%%%%%%%%%%%%%%%%%%%%%%%%%%%%%%%%%%%%%

As it was mentioned earlier, a large MCE also characterizes a distinctly different class of
materials, where the low temperature properties are governed by pronounced quantum many-body
effects. These materials exhibit a field-induced QCPs (a zero-temperature phase
transition), and the MCE has been used to study their quantum criticality. The adiabatic
demagnetization curves of extended compass model in the transverse
field is presented in Figs. (\ref{fig6}) and (\ref{fig7}) for constant initial
value of magnetic field ($h_{i}=3$) and several
starting temperatures ($T_{i}$) for $J_{2}=2$ and $J_{2}=0.8$, respectively. The lowest
temperatures of an adiabatic process are reached at $h^{\ast}_{1}<h_{c_{2}}=h_{0}$ and 
$h^{\ast}_{2}>h_{c_{1}}=h_{\pi}$.
At extremely low temperatures the difference between the demagnetization fields ($h^{\ast}_{1}, h^{\ast}_{2}$)
and critical fields ($h_{c_{1}}, h_{c_{2}}$)
becomes very small.  Clearly, the QCM in transverse field chain cools down to
lower temperatures near the second critical field ($h_{0}$) than the one close to the first critical 
field ($h_{\pi}$).

However, the Gr\"{u}neisen ratio (cooling rate) is noteworthy quantity to specify the second-order
transitions since it necessarily diverges near QCPs and the divergent behavior obeys the universal scaling
law \cite{R},

\bea
\no
\Gamma_{h}(T\rightarrow0,h)=-G_{h}\frac{1}{h-h_{c}},
\eea

where $G_{h}$ is a universal amplitude and the value $G_{h}=-1$ expected
for a $Z_{2}$-symmetry in the one dimension \cite{Zhitomirsky2}. Moreover, the sign of the Gr\"{u}neisen ratio
changes as entropy accumulates near a quantum critical point \cite{Sznajd}. The sign change
along with the divergence lead to strong signatures of the Gr\"{u}neisen parameter near QCPs.
The cooling rate of the extended quantum compass model is plotted versus the magnetic field
for $J_{2}=2$ and $J_{2}=0.9$ in Figs. (\ref{fig8}) and (\ref{fig9}) respectively.
As it can be seen from Fig.(\ref{fig1}), for $J_{2}=2$, curves show two quantum phase transition
at $h_{c_{1}}=h_{\pi}=1$ and $h_{c_{2}}=h_{0}=\sqrt{3}$ and for $J_{2}=0.9$ show just
one quantum phase transition at $h_{c}=h_{0}=\sqrt{1.9}$.
For low temperature, the quantum phase transition at $h_{c_{1}}=h_{\pi}=1$ and $h_{c_{2}}=h_{0}=\sqrt{3}$ are signaled
by sign changes of the cooling rate $\Gamma_{h}$ from negative to positive values upon increasing field,
see Fig.(\ref{fig8}). However in Fig.(\ref{fig9}) the cooling rate $\Gamma_{h}$ changes the sign when the
magnetic field crosses the QCP ($h_{c}=h_{0}=\sqrt{1.9}$). By increasing the temperature the magnitude of
the peaks reduce and all features are washed out which implies that thermal fluctuations are already strong enough
to drive the system to the excited state where no quantum phase transition can be seen.
In other words, the strong enhancement of the magnetocaloric effect arising
from quantum fluctuations near a $h$-induced quantum-critical point can be used for finding
an efficient and flexible high performance magnetic cooling over an extended
temperature range.

\section{Extended Quantum compass Model ($h_{1}=h_{2}=0$)\label{ED}}

%%%%%%%%%%%%%%%%%%%%%%  Fig.10   %%%%%%%%%%%%%%%%%%%%%%%
\begin{figure}
\begin{center}
\includegraphics[width=9.5cm]{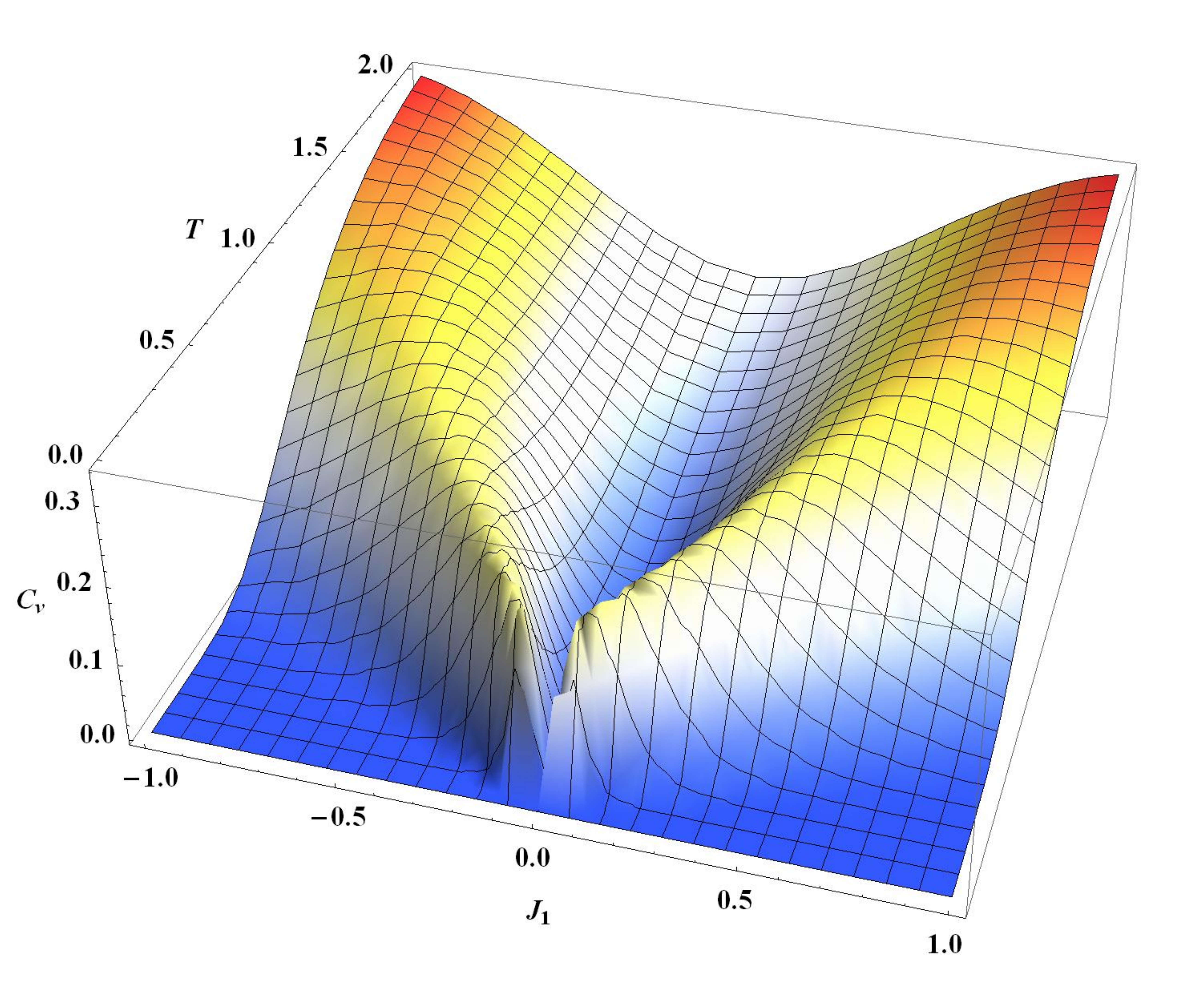}
\caption{(Color online) The three-dimensional plot of specific heat of
extended quantum compass model versus temperature and the parameter which
derives the first-order transition ($J_{1}$) for $J_{2}=2$.} \label{fig10}
\end{center}
\end{figure}
%%%%%%%%%%%%%%%%%%%%%%%%%%%%%%%%%%%%%%%%%%%%%%%%%%%%%%%

%%%%%%%%%%%%%%%%%%%%%%  Fig.11  %%%%%%%%%%%%%%%%%%%%%%%
\begin{figure}
\begin{center}
\includegraphics[width=9.5cm]{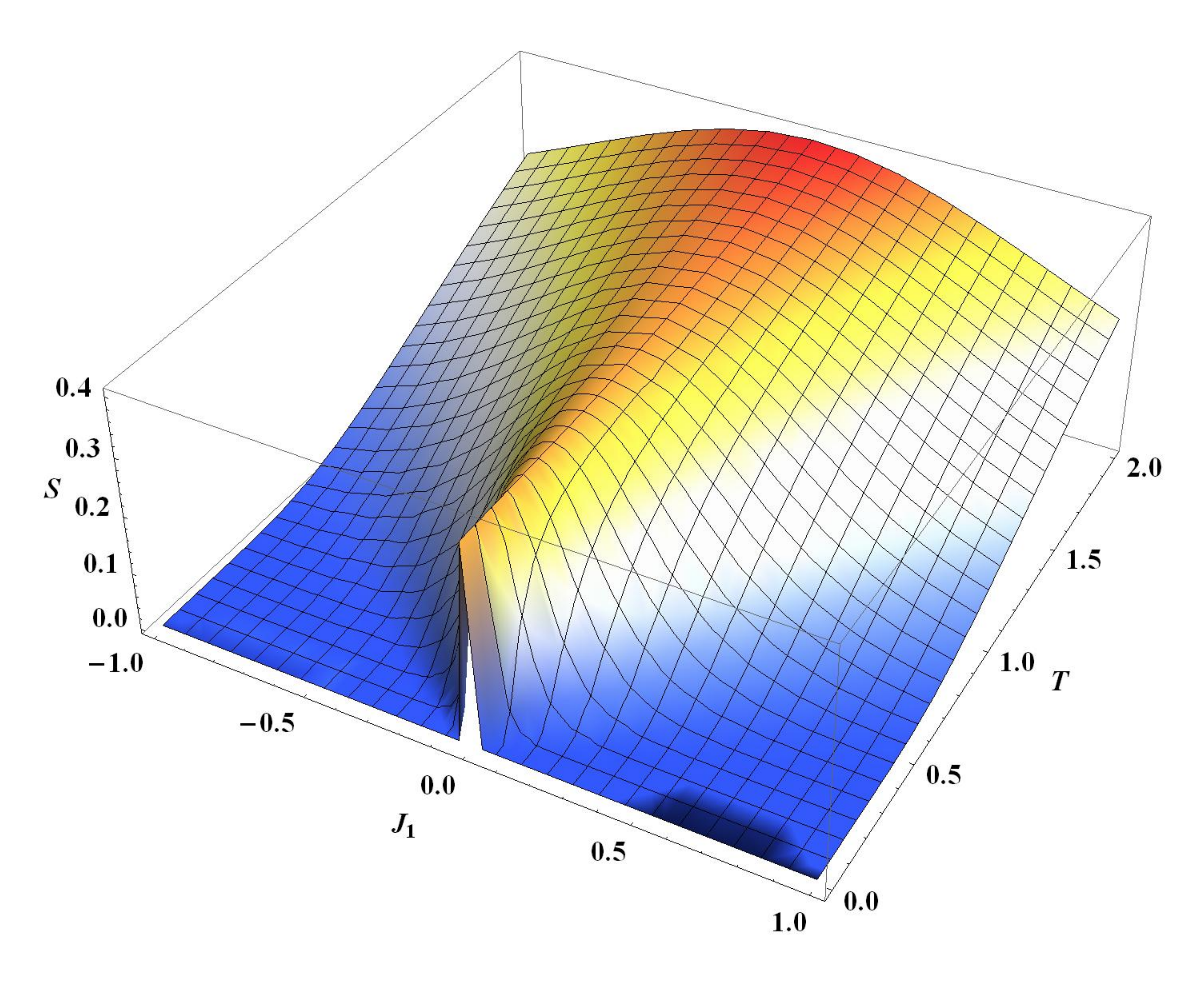}
\caption{(Color online) The three-dimensional panorama of entropy of
extended quantum compass model versus temperature and $J_{1}$ for $J_{2}=2$.} \label{fig11}
\end{center}
\end{figure}
%%%%%%%%%%%%%%%%%%%%%%%%%%%%%%%%%%%%%%%%%%%%%%%%%%%%%%%

%%%%%%%%%%%%%%%%%%%%%%  Fig.12   %%%%%%%%%%%%%%%%%%%%%%%
\begin{figure}
\begin{center}
\includegraphics[width=9cm]{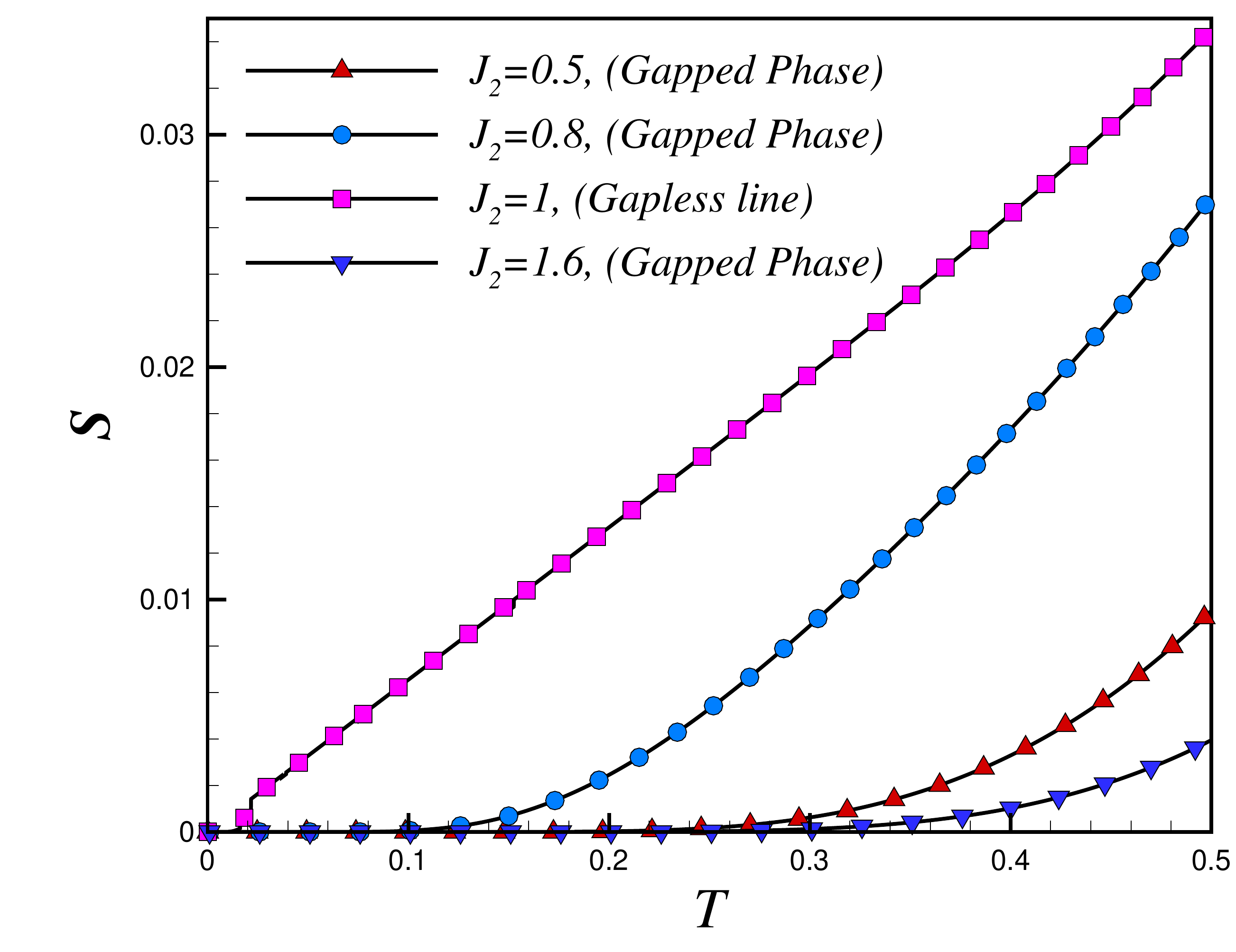}
\caption{(Color online) Entropy of the extended quantum compass model
versus $T$ for $J_{2}=0.5$, $J_{2}=0.8$, $J_{2}=1$ and $J_{2}=1.6$. Entropy on the
gapless line ($J_{2}=1$) shows the linear behavior.} \label{fig12}
\end{center}
\end{figure}
%%%%%%%%%%%%%%%%%%%%%%%%%%%%%%%%%%%%%%%%%%%%%%%%%%%%%%%

The complete phase diagram of the extended compass model has been reported in
Refs. [\onlinecite{Eriksson}] and [\onlinecite{Mahdavifar}]. They have shown that this model is always gapful
except at the critical line. The multicritical point located on the intersection of the first-order ($J_{1}/L_{1}=0$) 
and second order transition ($J_{2}/L_{1}=1$) lines (For simplicity we take $L_{1}=1$).

The specific heat of extended quantum compass model versus temperature ($T$) and the parameter 
deriving the first-order transition ($J_{1}$) is plotted in Fig. (\ref{fig10}). As it is clear, the
minimum of the specific heat occurs on the first order transition line ($J_{1}=0$) for any arbitrary
temperature. Fig. (\ref{fig11}) shows the three-dimensional panorama of the entropy versus $T$ and $J_{1}$
which manifests that the maximum of the entropy lies on the first order transition line ($J_{1}=0$).
It is worthy to mention that the system is gapful on this line where the degenerate
ground-state separates from the excited state.
So, the prominent characteristic of the specific heat and entropy on the first-order transition line inherited from
the existence of a gap.

\begin{figure}
\begin{center}
\includegraphics[width=9.2cm]{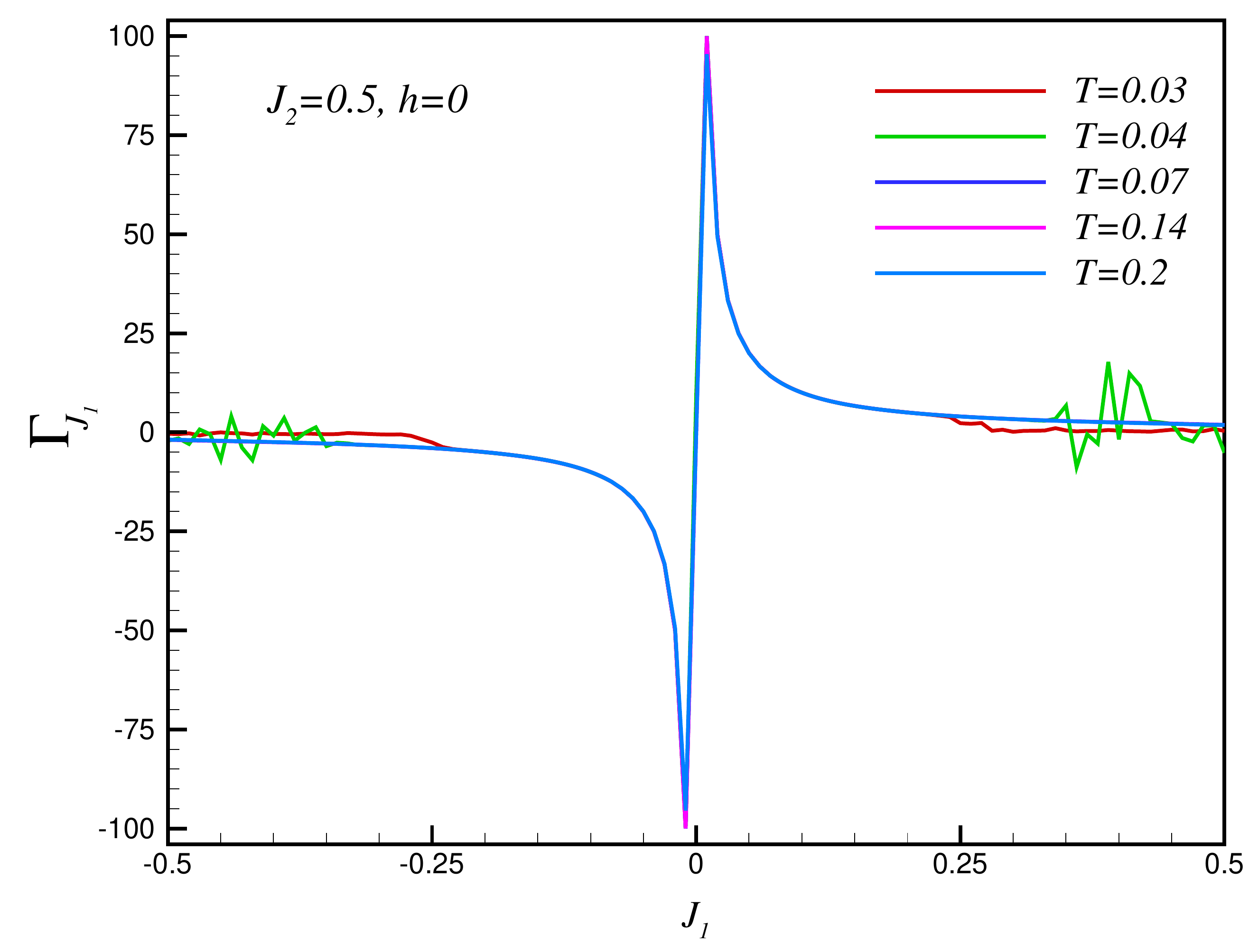}
\caption{(Color online) Cooling rate ($\Gamma_{J_{1}}$) versus
$J_{1}$ for different values of temperature, $J_{2}=2$.} \label{fig13}
\end{center}
\end{figure}

\begin{figure}
\begin{center}
\includegraphics[width=9cm]{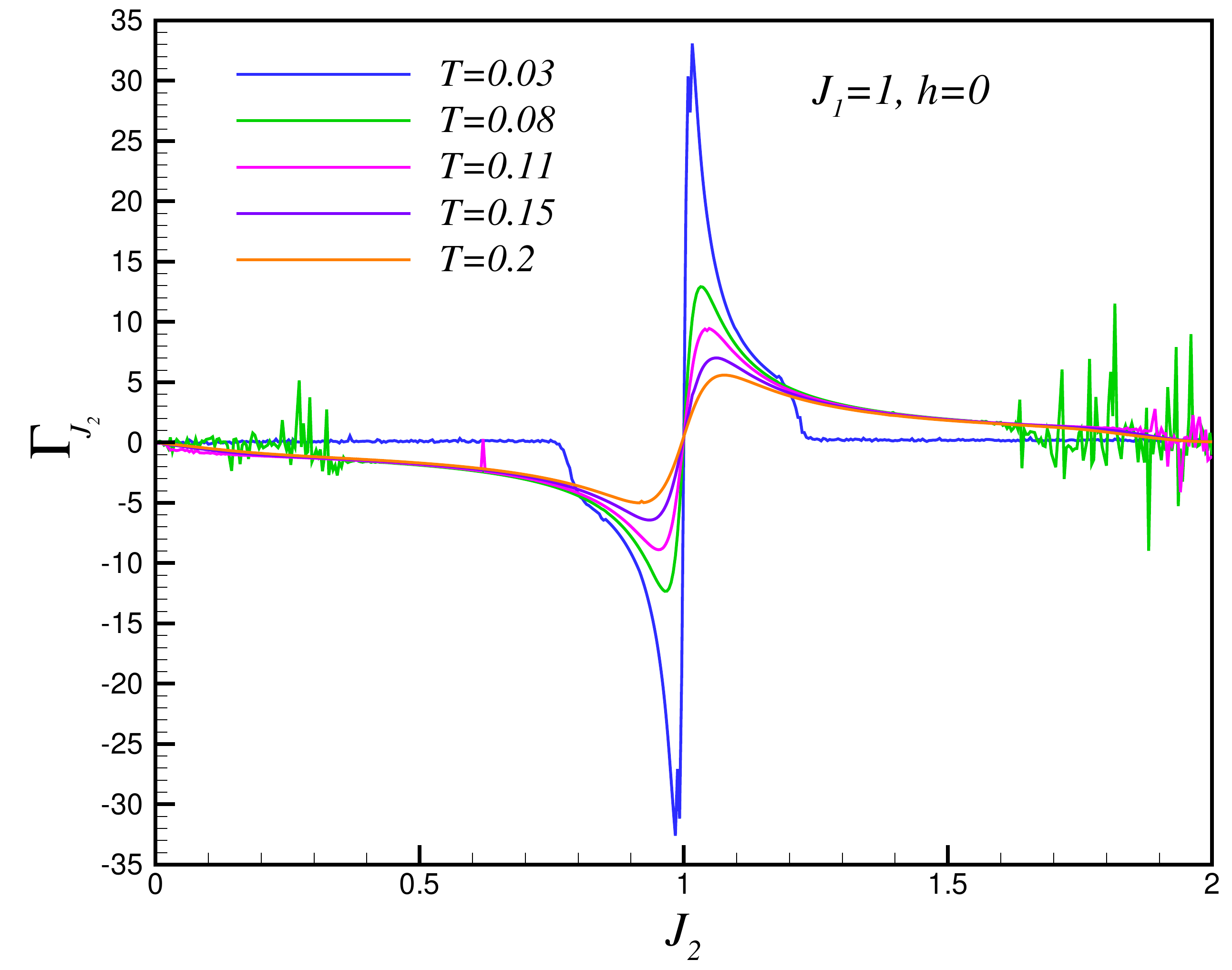}
\caption{(Color online) The variations of cooling rate ($\Gamma_{J_{2}}$) with
the parameter that derives the second-order transition ($J_{2}$) for different
values of temperature, $J_{1}=1$.} \label{fig14}
\end{center}
\end{figure}

However, the maximum entropy and specific heat falls out from the second order transition line ($J_{2}=1$)
for extremely low temperatures. On this line the system is gapless and entropy is linear in $T$ ($S\propto T$)
for low temperatures while in the gaped cases ($J_{2}\neq 1$) we expect activated behavior \cite{Trippe}
\bea
\no
S\propto \exp(-\Delta/T)
\eea

where $\Delta$ is gap in the excitation spectrum. The low temperature behavior of the entropy is plotted in
Fig. (\ref{fig12}) which verifies the linear and exponential behaviors of the entropy on the transition line
and gapped phases. The asymptotic behavior is $S\rightarrow0$ for $T\rightarrow0$ in all cases.
The cooling rate is plotted versus $J_{1}$ and $J_{2}$ in Figs. (\ref{fig13})
and (\ref{fig14}), respectively.

%%%%%%%%%%%%%%%%%%%%%%  Fig.13   %%%%%%%%%%%%%%%%%%%%%%%
%%%%%%%%%%%%%%%%%%%%%%%%%%%%%%%%%%%%%%%%%%%%%%%%%%%%%%%

%%%%%%%%%%%%%%%%%%%%%%  Fig.14   %%%%%%%%%%%%%%%%%%%%%%%
%%%%%%%%%%%%%%%%%%%%%%%%%%%%%%%%%%%%%%%%%%%%%%%%%%%%%%%

The cooling rate dependence on $J_{1}$ is plotted in
Fig. (\ref{fig13}) for different values of temperature. It is shown that the first-order
transitions are signaled by very sharp and pronounced positive and negative peaks at the
transition point ($J_{1}=0$). Further the only
difference between the curves, which correspond to different temperatures, is difference in the strength 
of the peaks.
Fig. (\ref{fig14}) shows the variations of cooling rate with $J_{2}$. The QCP pinpointed by sign changes of
the cooling rate $\Gamma_{J_{2}}$ from negative to positive values upon increasing $J_{2}$. The magnitude of
the peaks grows rapidly with decreasing the temperature.

\section{Summary and conclusions \label{conclusion}}
In this paper we have studied the thermodynamic properties of
the one dimensional extended quantum compass model in peresence/absent of transverse field.
We have presented the specific heat, entropy, adiabatic demagnetization curves and Gr\"{u}neisen parameter,
which is proportional to the magnetocaloric effect, as a function of the external magnetic
field on the thermodynamic limit and at finite temperatures.
We have used the exact result for the entropy to illustrate
that field-induced quantum phase transitions give
rise to maxima of the low-temperature entropy, or equivalently
minima of the isentropes. This leads to cooling
during adiabatic (de)magnetization processes where the
lowest temperature is reached close to the quantum phase
transition. As a consequence, we found a large
positive (negative) values of the normalized cooling rate
for magnetic fields slightly above (below) the critical
fields.
The general features of the entropy should not depend
on the specific choice of the magnetic field $h$ as control
parameter and indeed similar behavior is found as a function
of the exchange couplings.
The low-temperature asymptotics of the entropy $S$
is exponentially activated in the gapped phases and is linear
in $T$  on the second order transition line.

%%%%%%%%%%%%%%%%%%%%%%%%%%%%%%%%%%%%%%%%%%%%%%%%%%%%%%%%%%%%%%%%%
%\begin{acknowledgments}
%The authors would like to acknowledge for useful discussions and comments.
%\end{acknowledgments}
%%%%%%%%%%%%%%%%%%%%%%%%%%%%%%%%%%%%%%%%%%%%%%%%%%%%%%%%%%%%%%%%

\end{document}